\documentclass[12pt]{article}
\usepackage{graphicx}

\newsavebox{\sboxpubnumber}
\newsavebox{\sboxpubdate}
\newcommand{\pubdate}[1]{\begin{lrbox}{\sboxpubdate}{#1}\end{lrbox}}
\newcommand{\pubnumber}[1]{\begin{lrbox}{\sboxpubnumber}{\begin{tabular}{l} #1 \\
				 \usebox{\sboxpubdate}
				 \end{tabular}}
                           \end{lrbox}
                           \pubblock}
\newcommand{\mapright}[1]{\smash{\mathop{\hbox to 1cm{\rightarrowfill}}\limits_{#1}}}
\newcommand{\Title}[1]{\begin{center} {\Large #1 } \end{center}}
\newcommand{\Author}[1]{\begin{center}{ \sc #1} \end{center}}
\newcommand{\Address}[1]{\begin{center}{ \it #1} \end{center}}

\newcommand{\pubblock}{\rightline{
			\usebox{\sboxpubnumber}}}
\newenvironment{Abstract}{\begin{quotation}  }{\end{quotation}}
\newenvironment{Presented}{\begin{quotation} \begin{center}
             PRESENTED AT\end{center}\bigskip
      \begin{center}\begin{large}}{\end{large}\end{center}
      \end{quotation}}
\newcommand{\Acknowledgements}{\bigskip  \bigskip \begin{center} \begin{large}
             \bf ACKNOWLEDGEMENTS \end{large}\end{center}}

\begin{document}

\begin{titlepage}
\pubdate{\today}                    
\pubnumber{} 

\vfill
\Title{Cosmological magnetic fields induced by metric perturbations
after inflation}
\vfill
\Author{Antonio L. Maroto\footnote{On leave of absence from Dept. F\'{\i}sica
Te\'orica, Universidad Complutense de Madrid, 28040, Madrid, Spain}}
\Address{Physics Department, Stanford University \\
         Stanford CA 94305-4060, USA}
\vfill
\vfill
\begin{Abstract}
We consider the amplification of electromagnetic quantum vacuum fluctuations
induced by the presence of metric perturbations at the end of inflation.
We obtain the amplitude of the corresponding  
magnetic fields on super-Hubble scales and compare
it  with the requirements of the galactic dynamo mechanism for different
values of the spectral index. Finally we discuss  the possible effects
of the dissipation of such  fields in the form of gravitational waves. 
\end{Abstract}
\vfill
\begin{Presented}
    COSMO-01 \\
    Rovaniemi, Finland, \\
    August 29 -- September 4, 2001
\end{Presented}
\vfill
\end{titlepage}
\def\thefootnote{\fnsymbol{footnote}}
\setcounter{footnote}{0}

\section{Introduction}
The origin of the large scale magnetic fields observed in galaxies and
galaxy clusters still remains an open problem in astrophysics 
(see \cite{Kronberg,Grasso} and references therein). Observations
show that  typical galactic fields  have  coherence lengths around
$1-10$ kpc and strengths of $1$ $\mu$G. Magnetic fields in galaxy clusters 
are
stronger $1-10$ $\mu$G, with larger coherence lengths $10-100$ kpc. 

Although their origin is unknown, there are certain theoretical limits
on their primordial strength. Thus, any homogeneous magnetic field 
existing before decoupling
should be weaker than  $B_0\leq 10^{-9}$ G  in order to avoid the 
production of excessive anisotropies in the cosmic microwave background 
radiation \cite{Grasso,Adams}.
On the other hand, more recently \cite{Caprini}, it has been shown 
that magnetic fields present before nucleosynthesis can be very 
efficiently dissipated in the form of gravity waves. The limits imposed
by nucleosynthesis on the maximum allowed additional energy density in the
form of gravity waves can be translated into limits on the primordial
strength of magnetic fields. Those limits can be as 
stringent as  $B_0\leq 10^{-39}$ G,
for magnetic field generated during inflation with a thermal spectrum.
   
Concerning the origin of the fields, there are several mechanisms proposed
in the literarature which can be roughly classified into two groups: 
first, magnetic fields can be produced by certain charge separation
mechanisms during galaxy formation \cite{Harrison}, 
or  second, by the amplification 
of preexisiting seed fields.  In the second case, the 
amplification  can be achieved either by    
the adiabatic compression of the fields in the collapse
of the protogalactic cloud, which requires a seed field of 
$B_0\sim 10^{-10}$ G; or by the galactic dynamo mechanism, where  
the differential rotation of the galaxy is able to transfer kinetic 
energy into magnetic field. In this last case, the limits on the
primordial seed fields at decoupling are 
in the range \cite{Davis}
$B_{dec}\sim 10^{-17}-10^{-20}$ G for a flat universe without
cosmological constant. For a flat universe with cosmological constant
the limits are relaxed up to $B_{dec}\sim 10^{-25}-10^{-30}$ G. 
Concerning the generation of the neccesary seeds, there are also two main types
of explanations. On one hand, those based on phase transitions in the early
universe, whose main difficulty is that the typical field coherence length
is very small. On the other hand, 
we have the amplification of electromagnetic (EM) quantum fluctuations
during inflation \cite{Turner}. The main problem in this case 
is that in order to get some amplification, it is necessary to break 
the conformal triviality
of Maxwell equations in Friedmann-Robertson-Walker (FRW) backgrounds.
Recently, we have  have  explored  this possibility 
in \cite{maroto}, and in the
present paper we  review our main results. 
Let us then start by studying Maxwell equations in a cosmological background.

\section{Maxwell equations in cosmological backgrounds}

Consider Maxwell equations
\begin{eqnarray}
\nabla_\mu F^{\mu\nu}=0,
\label{Maxwell}
\end{eqnarray}
in a Friedmann-Robertson-Walker background:
\begin{eqnarray}
g^{0}_{\mu\nu}dx^\mu dx^\nu=a^2 (\eta)(d\eta^2-\delta_{ij}dx^idx^j)
\end{eqnarray}
with the 
Coulomb gauge condition $\vec \nabla \cdot \vec A=0 \Rightarrow A_0=0$. 

The equation for the  Fourier modes  
$A_i(\vec q,\eta)$ is:
\begin{eqnarray}
\frac{d^2}{d\eta^2}A_i(\vec q,\eta)+q^2
A_i(\vec q,\eta)=0
\end{eqnarray} 
The trivial solutions are given by positive and negative 
frequency plane-waves: $A_i(\vec q,\eta)
\propto e^{\pm iq\eta}$ which are valid  $\forall \eta$.
This implies that if we start at $\eta\rightarrow -\infty$ 
with a positive frequency plane-wave, we will end up in 
$\eta\rightarrow +\infty$ with the same kind of solution, i.e. there is
no mixing between positive and negative frequency modes, and therefore
there is no photon production. This is known as the conformal triviality
of Maxwell equtions in FRW backgrounds. 

Therefore, in order to get magnetic field
amplification we need to break conformal invariance. There are 
several proposal
in the literature, which in general require certain modifications of 
Maxwell electromagnetism \cite{Turner,Dolgov} or the introduction of
 new fields  \cite{Ratra,giovan}.
 Here we explore the alternative possibility, i.e. we will not modify 
Maxwell electromagnetism but the background metric, including scalar
perturbations. This possibility is rather natural since we know that metric
perturbations are present in our universe. Let us then consider
the inhomogeneous background $g_{\mu\nu}=g^{0}_{\mu\nu}+h_{\mu\nu}$,
where:
\begin{eqnarray}
h_{\mu\nu}dx^\mu dx^\nu=2\Phi\; a^2(\eta)(d\eta^2+\delta_{ij}dx^i dx^j)
\nonumber
\end{eqnarray}
is the most general form of the linearized scalar metric perturbations in the
longitudinal gauge, where $\Phi(\eta,\vec x)$ is the  gauge-invariant 
gravitational potential.

Substituting this form of the metric into (\ref{Maxwell}),
we obtain the following linearized equations:
\begin{eqnarray}
\frac{\partial}{\partial x^i}\left((1-2\Phi)(\partial_i A_0-\partial_0
  A_i)
\right)&=&0, \nonumber\\ 
\frac{\partial}{\partial \eta}\left((1-2\Phi)(\partial_i A_0-\partial_0
  A_i)
\right)&+&  
\frac{\partial}{\partial x^j}\left((1+2\Phi)(\partial_j A_i-\partial_i
  A_j)
\right)=0
\end{eqnarray}
Now, because of the presence of the inhomogeneous perturbations, plane
waves are no longer exact solutions, and in principle we have the
possibility of mode mixing, and as
a consequence quantum vacuum fluctuations can be amplified. Particle
production from inhomogeneous sources has been considered also in
\cite{Frieman,noise} 

\section{Photon production}

In order to define the asymptotic
vacuum states, let us assume:
$\Phi\rightarrow 0$ when $\eta\rightarrow \pm \infty$. This is a
good approximation in the remote past, since  the own metric perturbations
are generated during inflation. It is also a
good approximation in the asymptotic future 
for perturbations reentering the horizon right after inflation,
since they  oscillate with damped amplitude. As we will see later, these
perturbations will give the leading contributions in our results. 
 
Consider positive frequency plane-wave solution in the $in$
region:
\begin{eqnarray}
A_\mu^{\vec k,\lambda}(x,in)
\mapright{\eta\rightarrow -\infty} 
A_\mu^{(0)\vec k,\lambda}(x)=\frac{1}{\sqrt{2kV}}
\epsilon_\mu(\vec k,\lambda)e^{i(\vec k\vec x-k\eta)}
\label{zero}
\end{eqnarray}
We are working in a box with finite comoving volume $V$ and the 
 two physical polarization states satisfy: $\vec\epsilon(\vec k,\lambda)
\cdot \vec k=0$, $\epsilon_0(\vec k,\lambda)=0$.

It is easy to see that in the $out$ region, the solution (\ref{zero}) 
will become a linear 
superposition of positive and negative frequency modes with
different polarizations and different momenta.
\begin{eqnarray}
A_\mu^{\vec k,\lambda}(x,in)\mapright{
\eta\rightarrow +\infty} 
\sum_{\lambda'}\sum_{q}\left(\alpha_{kq 
\lambda\lambda'}
\frac{\epsilon_\mu(\vec q,\lambda')}{\sqrt{2qV}}
e^{i(\vec q\vec x-q\eta)}+\beta_{kq \lambda\lambda'}
\frac{\epsilon_\mu^*(\vec q,\lambda')}{\sqrt{2qV}}
e^{-i(\vec q\vec x-q\eta)}\right)
\label{super}
\end{eqnarray}

If we quantize these modes, we find for the creation and
annihilation $out$ operators 
$a_{k\lambda}^{out\,\dagger}$, $a_{k\lambda}^{out}$ 
a similar expansion in terms of $in$ operators:
\begin{eqnarray}
a_{k\lambda}^{out}= 
\sum_{\lambda'}\sum_{q}\left(\alpha_{qk 
\lambda'\lambda}a_{q\lambda'}^{in}
 +\beta_{qk 
\lambda'\lambda}^*
a_{q\lambda'}^{in\;\dagger}\right)
\end{eqnarray}
This implies that if we started in the remote past in the
vacuum state $\vert 0, in\rangle$, then the number of particles
created in the $out$ region with momentum $k$ and
polarization $\lambda$ will be given by:
\begin{eqnarray}
\langle 0, in \vert N_{k\lambda}^{out} 
\vert 0,in\rangle
=\sum_{q}\sum_{\lambda'} \vert 
\beta_{qk\lambda\lambda'} \vert^2
\neq 0
\end{eqnarray}
In order to obtain the total number of photons created by the 
perturbations, we need to know the value of the Bogolyubov
coefficients $\beta_{qk\lambda\lambda'}$, for that purpose we have to 
solve the equations of motion. Thus, we look for perturbative 
solutions in $\Phi$:
\begin{eqnarray}
A_\mu^{\vec k,\lambda}(x)=A_\mu^{(0)\vec k,\lambda}(x)
+A_\mu^{(1)\vec k,\lambda}(x)+ ... 
\end{eqnarray}
For the  spatial equations in Fourier space
we get up to first order \cite{maroto}:
\begin{eqnarray}
\frac{d^2}{d\eta^2}{A_i^{(1)\vec k, \lambda}}(\vec q,\eta)+q^2
A_i^{(1)\vec k, \lambda}(\vec q,\eta)
-J_i^{\vec k, \lambda}(\vec q,\eta)=0
\end{eqnarray}
where:
\begin{eqnarray}
J_i^{\vec k, \lambda}(\vec q,\eta)&=&-\sqrt{\frac{2k}{V}}\left(\left(i
\Phi'(\vec k+\vec q,\eta)+\frac{k^2-\vec k \cdot \vec q}{k}
\Phi(\vec k+\vec q,\eta)\right) 
\epsilon_i(\vec k,\lambda) e^{-ik\eta} \right.\\ \nonumber
&+&\left.(\vec \epsilon(\vec k,\lambda)
\cdot \vec q) \;\Phi(\vec k+\vec q,\eta)\frac{k_i}{k}e^{-ik\eta}
-i
\frac{\vec \epsilon(\vec k,\lambda)
\cdot \vec q}{q^2}\frac{d}{d\eta}\left(\Phi(\vec k+\vec q,\eta)
e^{-ik\eta}\right)q_i\right)
\end{eqnarray}
It is relatively easy to solve these equations  up to first order 
in perturbations. The result is:
\begin{eqnarray}
A_i^{\vec k, \lambda}(\vec q,\eta)=\frac{\epsilon_i
(\vec k,\lambda)}{\sqrt{2kV}}
\delta(\vec q-\vec k)e^{-ik\eta}
+\frac{1}{q}\int_{\eta_0}^\eta 
J_i^{\vec k, \lambda}(\vec q,\eta')\sin(q(\eta-\eta'))d\eta'
\end{eqnarray}
Comparing this solution with the
expansion in (\ref{super}), we get for the Bogolyubov coefficients:
\begin{eqnarray}
\beta_{kq \lambda\lambda'}=\frac{-i}{\sqrt{2qV}}\int_{\eta_0}^{\eta_1} 
\vec \epsilon \;(\vec q,\lambda')\cdot \vec J^{\;\vec k, 
\lambda}(\vec q,\eta) 
e^{-iq\eta}d\eta
\end{eqnarray}
where $\eta_0$ denotes the initial time of inflation and $\eta_1$ the
present time.

In the inflationary cosmology, metric perturbations are generated
when quantum fluctuations become super-Hubble sized during inflation 
and reenter the horizon during radiation or matter dominated eras
as classical fluctuations. Therefore, we will consider only
the effect of those super-Hubble scalar perturbations, whose
evolution is given by \cite{brand}:
\begin{eqnarray}
\Phi(\vec k,\eta)=C_k\frac{1}{a}\frac{d}{d\eta}\left(\frac{1}{a}\int a^2
  d\eta\right)+ D_k\frac{a'}{a^3} \simeq C_k \,{\cal F}(\eta)
\end{eqnarray}
where the second term decreases during inflation and can be 
ignored.
Introducing this form for the perturbations, we can obtain an explicit
expression for the total number of photons created with 
momentum $k_G=2\pi/\lambda_G$ (corresponding to a galactic scale wavelength 
$\lambda_G\sim 10$ kpc), in terms of the power spectrum of
metric perturbations. Thus we find:
\begin{eqnarray}
N_{k_G}\simeq \frac{4(2\pi)^{3/2}}{3k_G}
\int dk \,{\cal P}_{\Phi}(k)
\label{number}
\end{eqnarray}
where we have used $k\eta\ll 1$, which is valid for
super-Hubble perturbations, and we have defined:
\begin{eqnarray}
{\cal P}_\Phi(k)=\frac{k^3 
\vert C_k\vert ^2}{2\pi^2 V}=
A_S^2\left(\frac{k}{k_C}\right)^{n-1}
\end{eqnarray}
with $\lambda_C\simeq 3000 \;\mbox{Mpc}$ and the 
COBE normalization $A_S\simeq 5 \cdot 10^{-5}$. We have 
assumed a simple power-law behaviour for the power spectrum.
Such behaviour should be valid up to some high-frequency cutoff 
$k_{max}$, corresponding to the perturbation with smallest 
wavelength produced during inflation. Typically that 
wavelength is related to size of the horizon at the
end of inflation, and is given by $k_{max}\sim a_IH_I$.

\section{Magnetic field generation}

In order to relate the number of photons created with the  
magnetic field strength,
we  notice that very long wavelength photons can be seen as
static electric or magnetic fields \cite{Turner}. Because of the high
conductivity of the universe during the radiation dominated era, the electric
field components are damped very fast, whereas magnetic flux is
conserved, i.e. $Ba^2=$const\footnote{The growth of conductivity 
during reheating could have some effects on the amplification, 
for a detailed discussion see \cite{maroto,giovan2}}. 

Thus, the energy density in a magnetic field mode $B_k$ will be given by:
\begin{eqnarray}
\rho_B(\omega)=\omega \frac{d\rho_B}{d\omega}=
\frac{\vert B_k
  \vert^2}{2}=w^4 N_k,
\end{eqnarray}
where $w=k/a$ is the physical wavenumber. Using the result for the 
occupation number in (\ref{number}), we obtain:
\begin{eqnarray}
\vert B_{k_G}^{dec}\vert \simeq 
\frac{2^{3/2}(2\pi)^{3/4}\;A_S}{\sqrt{3n} \;a_{dec}^2}\;\frac{k_{max}^{n/2}\; 
k_G^{3/2}}{k_C^{(n-1)/2}}
\end{eqnarray}
From this expression we see that the magnetic field
spectrum is thermal, i.e.,  $B_k\sim k^{3/2}$. In order to compare this
result with  observations and with the limits imposed by the galactic dynamo
mechanism, we have plotted in Fig.1 the magnetic field strength at decoupling
versus the cut-off frequency $k_{max}$, 
for different values of the spectral index $n$. In general, the results
are several orders of magnitude below the observed strengths. The
dashed horizontal line represents the weakest field required to seed a 
galactic dynamo, corresponding to a flat universe with cosmological constant
$\Omega_\Lambda=0.7$, $\Omega_m=0.3$ and $h=0.5$ \cite{Davis}.

In order to estimate the value of $k_{max}$ in typical models of inflation, 
let us assume $k_{max}\sim a_I H_I$, where the $I$ subindex denotes the end
of inflation. For a model with $H_I=10^{13}$ GeV, we can estimate $a_I$ 
assuming that reheating is very efficient, so that all the energy density
in the inflaton field is instantaneously converted into radiation with
a temperature $T_R$. In this case, we have: 
\begin{eqnarray}
k_{max}\sim a_I H_I\sim 
\frac{a_{dec}T_{dec}H_I}{T_{RH}}
\end{eqnarray}
where we have assumed adiabatic evolution of the universe after
reheating. For a typical reheating temperature $T_R\sim 10^{15}$ GeV, we have
$k_{max}/k_C\simeq 10^{26}$, the corresponding
magnetic field from Fig. 1  is below the requirements of the
galactic dynamo.  Only very large values
of the cutoff frequency can give rise to a sufficiently strong seed magnetic
field. 
\begin{figure}[htb]
    \centering
    \includegraphics[height=8cm]{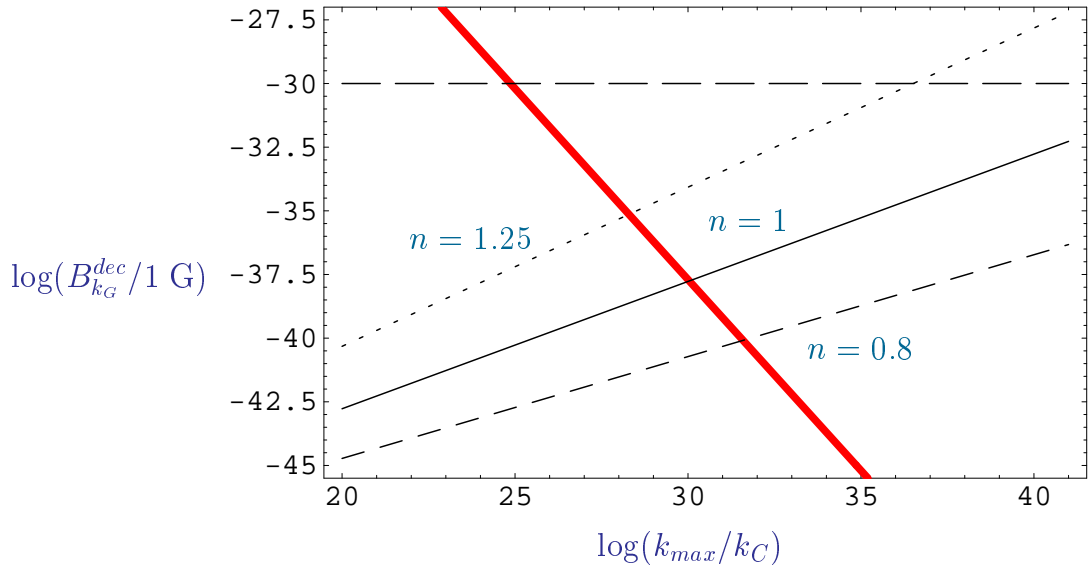}
    \caption{Magnetic field strength at decoupling on a galactic scale $k_G$
versus cut-off frequency $k_{max}$ for different values
of the spectral index $n$. The continuous line corresponds to the
scale-invariant Harrison-Zeldovich spectrum with $n=1$, the dashed line to
$n=0.8$ and the dotted one to $n=1.25$. The dashed horizontal line
corresponds to the weakest field required to seed a galactic dynamo, and
the thick line is the limit imposed by gravity waves production given by
 (\ref{gw}).}
    \label{fig:cosmo}
\end{figure}

\section{Damping of magnetic field into gravity waves}

Finally we will comment on the possible dissipation of magnetic fields in
the form of gravity waves \cite{Caprini}. It is well-known that the interaction
with the cosmic plasma is responsible for the damping of the magnetic field
at small scales due to viscosity. However, if magnetic field are produced at
very early times, the main dissipation mechanism is in the form of
gravity waves \cite{Caprini}, which are
not damped by viscosity since they couple very weakly to
matter. In that work, it has been shown that, in order
to avoid an excess of energy density in gravity waves,  that could  
modify the expansion rate
of the universe during nucleosyntheis, the strength of any stochastic 
magnetic field created before nucleosynthesis should satisfy:
\begin{eqnarray}
B_k<7\cdot 10^{-7} (\eta_{in} k)^{(n_B+3)/2} h_0\, {\cal N}(n_B)\;{\mbox G}
\label{gw}
\end{eqnarray}
where $B_k$ is the magnetic field mode strength (rescaled today)
 with momentum $k$, ${\cal N}(n_B)\sim 1$, $\eta_{in}=1/k_{max}$ and
  $n_B$ is the magnetic field spectral
index defined as $B_k^2\propto k^{n_B+3}$.  Notice that for a thermal spectrum 
$n_B=0$. In Fig. 1, we have plotted the above limit (thick line) 
as a function of $k_{max}$ for the magnetic fields produced by 
metric perturbations whose spectral index is thermal, 
as shown before. We see that, such limit  excludes the region 
of parameter space that could seed the galactic dynamo. However, still
we can use these bounds to impose some limits on the primordial spectrum of 
metric perturbations at small scales. Thus, in order to avoid excessive
gravity waves production (via magnetic fields), 
the cut-off frequency in the metric
perturbations power spectrum should be 
$k_{max}/k_C<10^{30}$ for a scalar spectral index $n\simeq 1$.

\section{Conclusions}

In this paper, we have reviewed some of the results obtained in \cite{maroto}
concerning the production of large-scale magnetic fields from metric
perturbations.  We have shown how the breaking of conformal invariance
induced by the presence of metric perturbations is able to produce
photons at the end of inflation, and we have related the occupation
number with the power spectrum of the metric perturbations.

We have compared the magnetic fields produced by this mechanism with 
the observations in galaxies and galaxy clusters and we have concluded that
they are several order of magnitude weaker. Even with the assistance of 
galactic dynamo mechanism, only for extreme values of the parameters, this
mechanism could explain the observations.

Finally, we have considered the dissipation of those magnetic fields in the
form of gravity waves. Following the results in \cite{Caprini}, we have shown 
that, like most of the  models of magnetogenesis before nucleosynthesis, the
one presented here would be excluded because of the  excessive 
production of gravity waves. 
However, we can use the same limits on
the primordial magnetic fields to impose certain bounds on the primordial
spectrum of metric perturbations at small scales. 
 
\Acknowledgements
This work has been partially supported by the CICYT (Spain) projects
AEN97-1693 and FPA2000-0956. The author  also acknowledges support from 
the Universidad Complutense del Amo Fellowship.

\end{document}